\documentclass[sort&compress
    ,final            
    ,numberedheadings 
    ,cmfonts          
  ]
  {aipproc}

\usepackage{subfigure}
\newcommand{\ce}[1]{Eq.~(\ref{#1})}
\newcommand{\be}{\begin{equation}}
\newcommand{\ee}{\end{equation}}
\newcommand{\bea}{\begin{eqnarray}}
\newcommand{\eea}{\end{eqnarray}}

\layoutstyle{6x9}

\begin{document}

\title{Two-photon decay of pseudoscalar quarkonia}

\author{J.P. Lansberg}{
  address={Institut f\"ur Theoretische Physik, Universit\"at Heidelberg, 69120 Heidelberg, Germany}
}
\author{T. N. Pham}{
  address={Centre de Physique Th\'{e}orique, CNRS,
Ecole Polytechnique, 91128 Palaiseau, France \\
E-mail: lansberg@tphys.uni-heidelberg.de,pham@cpht.polytechnique.fr}
}

\begin{abstract}

We report on our recent evaluation of  the two-photon width of the pseudoscalar quarkonia, $\eta_c(nS)$
 and  $\eta_b(nS)$ in an approach based on Heavy-Quark Spin Symmetry (HQSS). To what concerns the $1S$ 
state $\eta_c$, our parameter-free computation agrees with experiments, as well as most of other 
theoretical works. On the other hand, our computation for the $2S$-state looks $2S$ like a confirmation 
that there may exist an anomaly related to the decay of $\eta_c'$, especially in the light 
of the new preliminary result of the Belle collaboration. We also point out that the essentially
 model-independent ratio of $\eta_b$ two-photon width to the $\Upsilon$ leptonic width and 
the $\eta_b $ two-photon width could be used to extract the strong coupling constant $\alpha_s$. 

\end{abstract}

\maketitle

\renewcommand{\thefootnote}{\fnsymbol{footnote}}
\footnotetext{Presented by T.N. Pham at the Joint Meeting Heidelberg-Li\`ege-Paris-Wroclaw (HLPW08), 6-8 March, 2008, Spa, Belgium.}
\renewcommand{\thefootnote}{\arabic{footnote}}

\section{Introduction}
Since the $J/\psi$ discovery,  bound states of heavy quarks --the heavy quarkonia--  are expected 
to provide physicists with ideal means to study the main properties of QCD. 
With the time though, it appeared progressively that such quark systems are not so easy to understand and controversies
about their production  mechanisms are still going on~\cite{Brambilla,Lansberg:2006dh}.
Fortunately, the  physics involved in their decay seems to be rather well 
understood within the conventional framework of QCD~\cite{Brambilla,Colangelo}.
 
 However, recently two experimental estimations of the two-photon width of the $\eta'_c$, one published
by the CLEO collaboration \cite{Asner} ($\Gamma_{\gamma \gamma}(\eta'_c)= 1.3
\pm 0.6$ keV) and  another preliminary  by the Belle collaboration ($\Gamma_{\gamma \gamma}(\eta'_c)= 0.59
\pm 0.13 \pm 0.14$ keV)\cite{Nakazawa,eidelman} contradict most of the existing theoretical 
predictions lying in the range $3.7$ to $5.7$
keV~\cite{Ackleh,Kim,Ahmady,Munz,Chao,Ebert,Gerasimov,Crater}.
This is rather surprising since such electromagnetic decays of non-relativistic states should be 
rather easy to describe from a theoretical point of view.

It was therefore our purpose in~\cite{Lansberg:2006dw,Lansberg:2006sy} to reanalyse 
such a process from a very basic starting point: the heavy-quark spin symmetry (HQSS). Indeed, 
in a non-relativistic system, the difference between the $1S$ and $2S$ widths
would appear only at the level of the wave function at the origin. In virtue of such a symmetry,
 both wave functions should not differ much from the ones of $^3S_1$ state, which are a priori well known
from the leptonic decays of $J/\psi$ and $\psi'$. 

We report here on our effective approach based on HQSS and compare its results with other available 
calculations. Since none of them is able to predict both experimental measurements for
$\Gamma_{\gamma\gamma}(\eta_c)$ and $\Gamma_{\gamma\gamma}(\eta'_c)$, we also discuss some of the 
hypotheses made by CLEO and Belle for the extraction of those widths from their
experimental observables. We finally report on the corresponding predictions for $\eta_b(nS)$ states.

\section{Effective Lagrangian for $^{1}S_{0}$ Decay into two photons}

\begin{figure}[h!]
\centering
\includegraphics[height=4cm]{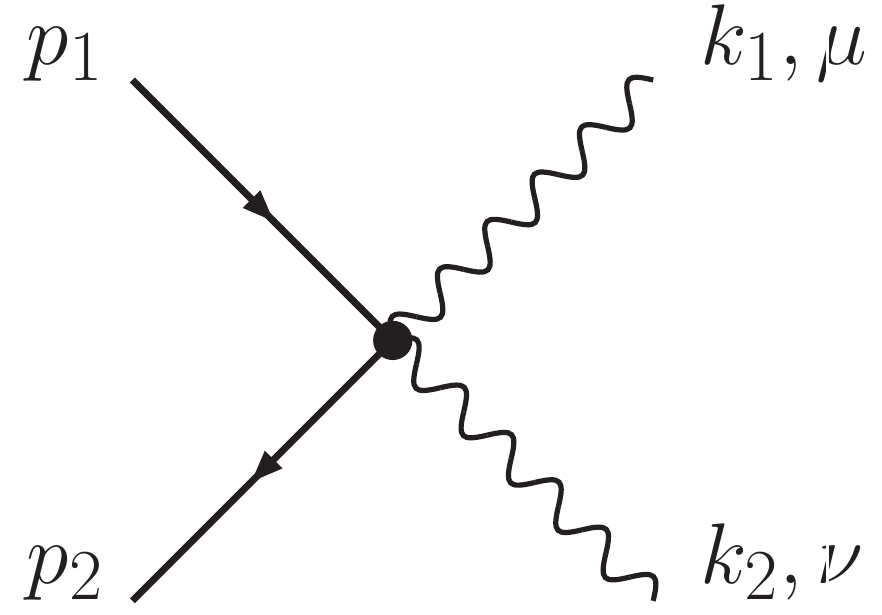}\quad
\includegraphics[height=4cm]{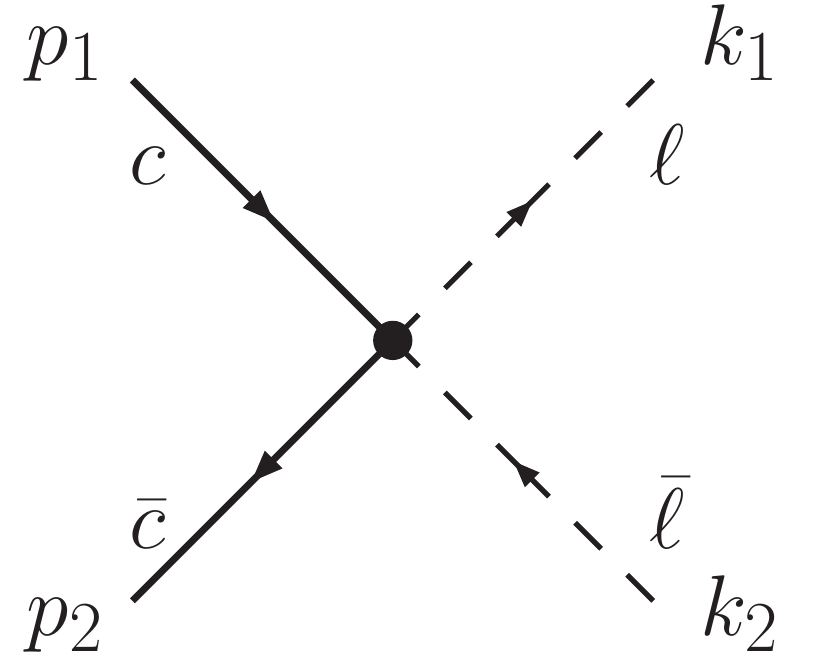}
\caption{Effective coupling between a $Q\bar{Q}$ and two photons (left)
  and a lepton pair (right)}
\label{Fig:decay}
\end{figure}

In the two-photon decay of a heavy quarkonium bound state, the 
outgoing-photon momentum is large compared to the relative momentum 
of the quark-antiquark bound state, which is ${\cal O}(\Lambda/m_{Q}) $, with
$\Lambda \ll m_{Q}$. One obtains an effective Lagrangian for the process 
$Q\bar{Q}\to \gamma\gamma$ (represented by the first diagram in 
Fig.\ref{Fig:decay}) by expanding the heavy-quark propagator
 in powers of $q^{2}/m_{Q}^{2}$, and  neglecting 
${\cal O}(q^{2}/m_{Q}^{2})$ terms ($q=p_{Q}-p_{\bar{Q}}$). Like leptonic
decay, the two-photon decay, in this approximation, is described  
by the following effective Lagrangian:

\bea
&&{\cal L}^{\gamma \gamma}_{\rm eff}=-i c_1(\bar{Q}\gamma^{\sigma}\gamma_{5} Q) \varepsilon_{\mu \nu \rho \sigma} F^{\mu\nu}
A^\rho \nonumber \\
&&{\cal L}^{\ell \bar \ell}_{\rm eff}=-  c_2(\bar{Q} \gamma^{\mu} Q) (\ell \gamma_\mu \bar \ell)
\label{Leff}
\eea
with 
\be
c_1\simeq \frac{Q_Q^2 (4\pi
  \alpha_{\rm em})}{M_{^1S_0}^2+b_{^1S_0} M_{^1S_0}} , \quad \quad 
c_2=\frac{Q_Q (4\pi \alpha_{\rm em})}{M_{^3S_1}^2} \,\,\, .
\label{coeff}
\ee

The factor $1/(M_{^1S_0}^2+b_{^1S_0} M_{^1S_0})$ in $c_{1}$ 
contains the binding-energy effects (the binding-energy $b$ is defined as
$b =2m_{Q} -M$) and is obtained  from the 
denominator of the heavy-quark 
propagator ($k_{1}, k_{2}$  being the outgoing-photon momenta):
\be
\frac{1}{[(k_{1}-k_{2})^{2}/4 - m_{Q}^{2}]} \,\,\, .
\ee
 The  decay amplitude  is then 
given by the matrix element of the axial-vector current
$\bar{Q}\gamma^{\mu}\gamma_{5}Q$ similar to the quarkonium leptonic
decay amplitude which is given by  the vector-current matrix element 
$\bar{Q}\gamma^{\mu}Q$ for $^3S_1 \to \ell^{+}\ell^{-}$. Thus for decays of 
 $S$-wave quarkonium into two photons or   a dilepton pair 
$\ell \bar \ell$, we have:
\bea
&&{\cal M}_{\ell \bar \ell}= 
Q_Q (4 \pi \alpha_{\rm em})\frac{f_{^3S_1}}{M_{^3S_1}} \varepsilon_\mu (\ell
\gamma^\mu \bar \ell) \nonumber\\
&&{\cal M}_{\gamma \gamma}=- 4 i Q_Q^2 (4 \pi \alpha_{\rm em})\frac{f_{^1S_0}}{M^2_{^1S_0}+b_{^1S_0} M_{^1S_0}} \varepsilon_{\mu \nu \rho \sigma} 
\varepsilon_1^\mu \varepsilon_2^\nu k_1^\rho  k_2^\sigma
\label{amp}
\eea
where  
\be
 \langle0|\bar Q \gamma^\mu  Q| ^3S_1 \rangle= f_{^3S_1}M_{^3S_1}\varepsilon^\mu, \quad
 \langle0|\bar Q \gamma^\mu \gamma_{5} Q| ^1S_0 \rangle= i f_{^1S_0}\,P^{\mu}.
\label{dc}
\ee
from which the decay rates are:
\be
\Gamma_{\ell \bar \ell}(^3S_1)=\frac{4 \pi Q_Q^2 \alpha^2_{\rm em}
  f_{^3S_1}^2}{3 M_{^3S_1}}, \qquad
\Gamma_{\gamma \gamma}(^1S_0)=\frac{4 \pi Q_Q^4 \alpha^2_{em} 
f_{^1S_0}^2}{M_{^1S_0}} .
\label{rate}
\ee
   By taking  $M_{^3S_1}f_{^3S_1}^2=12 |\psi(0)|^2$, we recover the usual
non-relativistic expression for the decay rate which, whose NLO QCD
 radiative corrections are given by 
\bea
&&\Gamma^{NLO}(^3S_1)= \Gamma^{LO}(^3S_1) \left(1- \frac{\alpha_s}{\pi}\frac{16}{3}\right) \\ 
&&\Gamma^{NLO}(^1S_0)= \Gamma^{LO}(^1S_0) \left(1- \frac{\alpha_s}{\pi}\frac{(20-\pi^2)}{3}\right)\,\,\, .
\label{rate1}
\eea

\section{Matrix elements of Local operators }

We have shown that in the  approximation of neglecting 
${\cal O}(q^{2}/m_{Q}^{2})$ terms, the two-photon
decay amplitude is given by the $^1S_0$ decay constant
$f_{^1S_0}$. We now  
derive a symmetry relation between $f_{^1S_0}$ and  
$f_{^3S_1}$, the  $^3S_1$ leptonic decay constant using the relativistic 
spin projection operators 
for a relativistic Bethe-Salpeter quarkonium bound state \cite{Pham}.
 
Consider now the matrix elements of local operators in a
fermion-antifermion system with a given spin $S$ and orbital angular 
momentum $L$ \cite{Kuhn,Guberina} :
\be
{\cal A} = \int
\frac{d^{4}\,q}{(2\,\pi)^{4}}\,{\rm Tr}\,{\cal O}(0)\chi(P,q)
\label{A1}
\ee
$P$ is the total 4-momentum of the quarkonium system, $q$ is the relative
4-momentum between the quark and anti-quark and $\chi(P,q) $ is
the Bethe-Salpeter wave function.
For a quarkonium system in  a  fixed total, orbital
and spin angular momentum,  $\chi(P,q) $ is given by (${\bf q}$ is the
relative 3-momentum vector of $q$)
\bea
\kern -0.5cm\chi(P,q; J,J_{z},L,S)\kern -0.2cm&& = \sum_{M,S_{z}}\,2\,\pi\delta(q^{0} - \frac{{\bf
    q}^{2}}{2\,m})\psi_{LM}({\bf q})\langle LM;SS_{z}|JJ_{z}\rangle \nonumber \\
\kern -0.5cm&& \times\sqrt{\frac{3}{m}}\sum_{s,\bar{s}}\,u(P/2 +q,s)\bar{v}(P/2-q,\bar{s})
\langle\frac{1}{2}s;\frac{1}{2}\bar{s}|S S_{z}\rangle \nonumber \\
\kern -0.5cm&&= \sum_{M,S_{z}}\, 2\,\pi\,\delta(q^{0} - \frac{{\bf
    q}^{2}}{2\,m})\psi_{LM}({\bf q}){\cal P}_{S S_{z}}(P,q)\langle LM;SS_{z}|JJ_{z}\rangle \,\,\, .
\label{chi}
\eea
The spin projection operators  ${\cal P}_{S S_{z}}(P,q) $ are 
\bea
&&\kern -0.5cm {\cal P}_{0,0}(P,q) = \sqrt{\frac{3}{8m^{3}}}[-(\rlap/P/2 +\rlap/q) +
m]\gamma_{5}\,[(\rlap/P/2 -\rlap/q) + m] \nonumber \\
&&\kern -0.5cm {\cal P}_{1,S_{z}}(P,q) = \sqrt{\frac{3}{8m^{3}}}[-(\rlap/P/2 +\rlap/q) +
m]\rlap/{\varepsilon}(S_{z})\,[(\rlap/P/2 -\rlap/q) + m]  \,\,\, .
\label{PJ}
\eea
  For $S$-wave quarkonium in a spin singlet $S=0$ and spin
triplet $S=1$ state:

\be
{\cal A}(^{2S +1}S_{J}) = {\rm Tr}\,({\cal O}(0)\,{\cal P}_{J\,J_{z}}(P,0))
\int \frac{d^{3}\,q}{(2\,\pi)^{3}}\,\psi_{00}(q) \,\,\, .
\label{A12}
\ee

In this expression  the $q$-dependence in the spin projection operator
has been dropped  and the integral 
in Eq.(\ref{A12}) is the $S$-wave function at the origin \cite{Guberina}:
\be
\int \frac{d^{3}\,q}{(2\,\pi)^{3}}\,\psi_{00}(q) =
\frac{1}{\sqrt{4\,\pi}}{\cal R}_{0}(0) \,\,\,\, .
\label{R0}
\ee

  Using Eq.(\ref{PJ}) and Eq.(\ref{A12}) to compute 
the matrix elements $\langle 0|\bar Q \gamma_\mu \gamma_{5} Q| P \rangle$
and $\langle 0|\bar Q \gamma_\mu  Q| V \rangle$ for the  singlet $S=0$
pseudo-scalar meson $P$ and for the  triplet $S=1$ vector meson $V$, 
we find, neglecting
quadratic ${\cal O}(q^{2})$ terms.
\be
f_{P} = \sqrt{\frac{3}{32\,\pi\,m^{3}}}\,{\cal R}_{0}(0)\, (4\,m) \ , \qquad
f_{V} = \sqrt{\frac{3}{32\,\pi\,m^{3}}}\,{\cal R}_{0}(0)\, \frac{(M^{2}
+ 4\,m^{2} )}{M}
\label{fPV}
\ee

Thus we get the relation 
\be 
f_{^1S_0}\simeq f_{^3S_1} + {\cal O} (b^2/M^2).
\ee
It is expected that this relation holds also for excited state of charmonium
and bottomonium where the binding terms ${\cal O}(b^{2}/M^{2})$ can be neglected.
This is a manifestation of heavy-quark spin symmetry(HQSS). In this limit,
the two-photon width of singlet $S=0$ quarkonium state can be obtained
from the leptonic width of triplet $S=0$ quarkonium state 
without using a bound state description.  This approach  differs 
from the traditional non-relativistic bound state approach  in
the use of local operators for which the matrix elements could be
measured or extracted from  physical quantities, or computed
from QCD sum rules \cite{Novikov,Reinders} and lattice QCD \cite{Dudek}.

   The ratio of the $\eta_{c}$ two-photon width
to $J/\psi$ leptonic width in the limit of HQSS is then:
\be
R_{\eta_{c}} = \frac{\Gamma_{\gamma \gamma}(\eta_c)}{\Gamma_{\ell \bar \ell}(J/\psi)}=3\,Q_c^2\,\frac{M_{J/\psi}}{M_{\eta_c}}\left(1+\frac{\alpha_s}{\pi}\frac{(\pi^{2}- 4)}{3}\right)\,\,\, .
\label{Rc}
\ee
For $M_{\eta_c}= M_{J/\psi} $, the above expression becomes the 
usual non-relativistic result \cite{Kwong, Pancheri} as mentioned above.
From  the measured $J/\psi$ leptonic width, we get
$\Gamma_{\gamma \gamma}(\eta_c)=7.46$ keV. Including  NLO QCD 
radiative corrections with $\alpha_{s}=0.26$, we find
$\Gamma_{\gamma \gamma}(\eta_c)=9.66$ keV
in near agreement  with the world average value $7.4 \pm 0.9 \pm 2.1$ keV.
A similar result is obtained in \cite{Pancheri} which gives 
$8.16 \pm 0.57 \pm 0.04$ keV .

Thus the  effective Lagrangian
approach  successfully predicts the $\eta_c$ two-photon width in
a simple, essentially model-independent manner. 

\section{HQSS predictions for $\Gamma_{\gamma \gamma}(\eta^{\prime}_c)$}

 To obtain the prediction for $\eta^{\prime}_c$, we shall apply HQSS
to $2S$ states. Thus, assuming  $f_{\eta^{\prime}_c}=f_{\psi^{\prime}}$, 
and neglecting binding-energy terms,  we find:
$\Gamma^{}_{\gamma \gamma}(\eta'_c)= \Gamma^{}_{\gamma \gamma}(\eta_c) 
\frac{f^2_{\psi'}}{f^2_{J/\psi}}=3.45~\hbox{keV}$. 
This value is  more than twice 
the evaluation by CLEO and five times the one by Belle. Our results is however
 nearly in agreement  with other theoretical calculations~\cite{Ackleh,Kim,Ahmady} as shown in
Table 1. Other approaches~\cite{Munz,Chao,Ebert,Crater} seem to be closer to the latter measurements but 
then undershoot clearly the measurements for $\eta_c$.

Including binding-energy terms,
for $M_{\eta_{c}}\simeq M_{J/\psi}$, 
$M_{\eta_{c}'}\simeq M_{\psi'}$, we have
\be
\Gamma_{\gamma \gamma}(\eta'_c)
=\Gamma_{\gamma \gamma}(\eta_c)
\left(\frac{1+b_{\eta_c}/ M_{\eta_c}}{1+b_{\eta'_c}/M_{\eta'_c}}\right)^2 
\times \left(\frac{\Gamma_{e^+e^-}(\psi')}{\Gamma_{e^+e^-}(J/\psi)}\right)
\label{be1}
\ee
which gives 
\be
\Gamma_{\gamma \gamma}(\eta'_c) = 4.1\,{\rm keV}\,\,\,\, .
\label{beeffect}
\ee

\begin{table}[h!]
\begin{tabular}{cccccccccccc}
\hline
$\Gamma_{\gamma \gamma}$ & This work &\cite{Ackleh} &
\cite{Kim}&\cite{Ahmady} &\cite{Munz}  &\cite{Chao} &\cite{Ebert}\kern -0.3cm&\cite{Crater}\\
\hline
$\eta_c$  & $7.5-10$ & $4.8$ & $7.14 \pm 0.95$&$11.8\pm 0.8\pm 0.6$&$3.5\pm 0.4$&$5.5$&$5.5$ &6.2 \\
$\eta'_c$ & $3.5-4.5$ & $3.7$&$4.44\pm 0.48$&$5.7\pm 0.5\pm 0.6$&
$1.38 \pm 0.3$ &$2.1$&$1.8$&3.36-1.95\\
\hline
\end{tabular}
\caption{  Theoretical predictions for 
$\Gamma_{\gamma \gamma}(\eta_c)$ and $\Gamma_{\gamma \gamma}(\eta'_c)$. (All values
are in units of keV).}\label{tab-res}
\end{table}

Binding-energy corrections seem then to worsen comparison with data and maybe point at anomaly
in the decay $\eta_c(nS)\to \gamma \gamma$. However, before drawing such a conclusion it is necessary
to discuss the experimental hypotheses made for the extraction of the aforementioned widths.

\section{${\cal B} (\eta_c(nS)\to KK\pi)$}

  The measured value from CLEO~\cite{Asner}~:
\bea
&&\Gamma_{\gamma \gamma}(\eta_{c}') = 1.3 \pm 0.6 \,{\rm keV},
\label{resCLEO}
\eea
was effectively done by
considering the following quantity :
\be
R(\eta_{c}'/\eta_{c}) = \frac{\Gamma_{\gamma \gamma}(\eta_{c}')\times
  {\cal B}(\eta_{c}' \to K_{S}K\pi)}{\Gamma_{\gamma \gamma}(\eta_{c})\times
  {\cal B}(\eta_{c} \to K_{S}K\pi)} = 0.18\pm0.05\pm 0.02.
\label{CLEO}
\ee

  To obtain  $\Gamma_{\gamma \gamma}(\eta_{c}') $ from
the above data, they made the assumption that 
\be
{\cal B}(\eta_{c}' \to K_{S}K\pi) \approx {\cal B}(\eta_{c} \to K_{S}K\pi)
\label{KKp}
\ee
and in turn found the result of~\ce{resCLEO}.

Such an assumption is in fact supported by a couple of observations.
Belle measurements of $B\to \eta_{c}K$ and 
$B\to \eta_{c}'K$ gives \cite{Choi}:
\be
R(\eta_{c}'K/\eta_{c}K)= \frac{{\cal B}(B^{0}\to \eta_{c}'K^{0})\times {\cal B}(\eta_{c}'\to
  K_{S}K^{+}\pi^{-})}{{\cal B}(B^{0}\to \eta_{c}K^{0})\times {\cal
    B}(\eta_{c}\to K_{S}K^{+}\pi^{-})} = 0.38\pm 0.12 \pm 0.05.
\label{Belle}
\ee

  Using the approximate equality~\ce{KKp}, one
would obtain
\be
\frac{{\cal B}(B^{0}\to \eta_{c}'K^{0})}{{\cal B}(B^{0}\to
  \eta_{c}K^{0}) } \approx 0.4,
\label{Belle1}
\ee
which agrees more or less with the QCD factorization (QCDF)
prediction \cite{Song}~:
\be
\frac{{\cal B}(B^{0}\to \eta_{c}'K^{0})}{{\cal B}(B^{0}\to
  \eta_{c}K^{0}) } \approx 0.9\times 
(\frac{f_{\eta_{c}'}}{f_{\eta_{c}}})^{2} \approx 0.45.
\label{qcdf}
\ee

On the other hand, it is expected from  $SU(2)$ flavor 
symmetry that one would have  the approximate equality
between the ratios  
\be 
\frac{{\cal B}(B^{0}\to \eta_{c}'K^{0})}{{\cal B}(B^{0}\to
  \eta_{c}K^{0})} \approx \frac{{\cal B}(B^{+}\to \eta_{c}'K^{+})}
{{\cal B}(B^{+}\to  \eta_{c}K^{+})}.
\ee 

This is indeed the case since the following ratio obtained 
from BABAR~\cite{Aubert:2005vi}
\be
\frac{{\cal B}(B^{+}\to \eta_{c}'K^{+})}
{{\cal B}(B^{+}\to  \eta_{c}K^{+})}  = 0.38\pm 0.25
\label{Babar}
\ee
corresponds again to \ce{Belle1}.

Those observations therefore tend to support  the assumption of
the approximate  equality between the $ \eta'_c \to KK\pi$ and 
$ \eta_c \to KK\pi$ branching ratio. This would confirm a 
small $\eta'_{c}\to \gamma\gamma$ decay rate as quoted above. 

Albeit, we also note that the good agreement with 
 QCDF predictions for the measured ratio 
${\cal B}(B^{0}\to \eta_{c}'K^{0})/{\cal B}(B^{0}\to  \eta_{c}K^{0}) $ and 
${\cal B}(B^{+}\to \eta_{c}'K^{+})/{\cal B}(B^{+}\to  \eta_{c}K^{+})$
at Belle and BABAR suggests that
$f_{\eta_{c}^{\prime}}/f_{\eta_{c}} \approx f_{\psi'}/f_{J/\psi}$,  
which in turn supports HQSS and our predicted value for the 
$\eta_{c}^{\prime}$ two-photon width which is  more than twice bigger
than the CLEO estimated value  shown above. More precisely, comparing 
$R(\eta_{c}'/\eta_{c})$  with $R(\eta_{c}'K/\eta_{c}K)$  and using QCDF value
given in Eq.(\ref{qcdf}), we find
\be
R(\eta_{c}'/\eta_{c}) \approx R(\eta_{c}'K/\eta_{c}K)/0.9 \,\,\,\, .
\label{Rt}
\ee
 The Belle data in \ce{Belle}  would then implies 
$R(\eta_{c}'/\eta_{c})\approx 0.42\pm 0.13\pm 0.05 $, twice
bigger than the CLEO data shown in \ce{CLEO}.

Since QCD gives ${\cal B}(\eta_{c}\to \gamma\gamma) \approx {\cal
  B}(\eta'_{c} \to \gamma\gamma) $ and the predicted ${\cal B}(\eta_{c}\to \gamma\gamma)$ agrees well with experiments, we expected that the large value for
the measured $\eta'_{c}$ total width would imply a large value
for the $\eta'_{c} $ two-photon width. Thus it is difficult 
to understand the very small recent Belle measured $\eta'_{c} $ two-photon width.
\section{ HQSS predictions for $\Gamma_{\gamma \gamma}(\eta_b)$ and
$\Gamma_{\gamma \gamma}(\eta'_b)$ }

  Since the $b$-quark mass is significantly
higher than the  $c$-quark mass, the effective Lagrangian and HQSS 
approach should work better for bottomonia decays  to leptons and photons.
We thus have:
\be
R_{\eta_{b}} = \frac{\Gamma_{\gamma \gamma}(\eta_b)}{\Gamma_{\ell \bar \ell}(\Upsilon)}=3\,Q_b^2\,\frac{M_{\Upsilon}}{M_{\eta_b}}\left(1+\frac{\alpha_s}{\pi}\frac{(\pi^{2}- 4)}{3}\right)
\label{Rb}
\ee
(neglecting the small $b_{\eta_{b}}/M_{\eta_b} $ binding-energy term).
This gives $\Gamma_{\gamma \gamma}(\eta_b)= 560\,\rm eV $ ($\alpha_s(M_{\Upsilon})=0.16$, $M_{\eta_b}=9300$ MeV) .

  For $\eta_{b}'$ and higher excited state, one has 
($M_{\eta_b} \simeq M_{\Upsilon}$ and $M_{\eta_{b}'}\simeq M_{\Upsilon'}$):
\be
\Gamma_{\gamma \gamma}(\eta'_b)
=\Gamma_{\gamma \gamma}(\eta_b)
\left(\frac{1+b_{\eta_b}/ M_{\eta_b}}{1+b_{\eta'_b}/M_{\eta'_b}}\right)^2 
 \left(\frac{\Gamma_{e^+e^-}(\Upsilon')}{\Gamma_{e^+e^-}(\Upsilon)}\right).
\label{etab}
\ee
which gives $\Gamma_{\gamma \gamma}(\eta'_b)= 250\,\rm eV $
and $\Gamma_{\gamma \gamma}(\eta''_b)= 187\,\rm eV $. In Table. \ref{tab-res1}
we give our prediction for the two-photon width of $\eta_{b}$, 
$\eta_{b}^{\prime} $ and $\eta''_{b}$  together with other  
theoretical predictions. We note that our predicted values are 
somewhat higher than other predicted values. 

  Eq.(\ref{Rb}) can be used to determine in a reliable way 
the value of $\alpha_{s}$. The momentum scale at which  
$\alpha_{s}$ is to be evaluated here  could be in principle be fixed 
with $R_{\eta_b}$. 

  Further check of consistency of the value for 
$\alpha_{s}$ may be possible in future measurements on the $\eta_{b}$
and its two-photon decay branching ratio: 
\be
 \frac{\Gamma_{\gamma\gamma}(\eta_b)}{\Gamma_{gg}(\eta_{b})}=
\frac{9}{2}\,Q_b^4\,\frac{\alpha^{2}_{em}}{\alpha^{2}_{s}}\left(1- 7.8\,\frac{\alpha_s}{\pi}\right) \,\,\, .
\label{br}
\ee
\begin{table}[h]
\begin{tabular}{cccccccccccc}
\hline
$\Gamma_{\gamma \gamma}$ & This work  &\cite{Schuler} &\cite{Lakhina}&\cite{Ackleh}& \cite{Kim}&
\cite{Ahmady} & \cite{Munz} &\cite{Ebert}&\cite{Godfrey}&\cite{Fabiano} \\
\hline
$\eta_b$ &   $560$  & $460$ & $230$&$170$& $384 \pm 47$& $520$ &$220 \pm 40$& $350$ &$214$&
$466\pm 101$\\
$\eta'_b$ &  $269$ & $200$& $70$& - & $191 \pm 25$& - & $110 \pm 20$& 150& 121&-\\
$\eta''_b$&  $208$ & -    & $40$ & - & - & -& $84 \pm 12$& 100& 90.6& -\\
\hline
\end{tabular}
\caption{ Summary of theoretical predictions for 
$\Gamma_{\gamma \gamma}(\eta_b)$, $\Gamma_{\gamma \gamma}(\eta'_b)$ and 
$\Gamma_{\gamma \gamma}(\eta''_b)$. (All values
are in units of eV).}\label{tab-res1}
\end{table}

\section{Conclusion}
 We have shown here that effective Lagrangian approach and HQSS 
can be used to compute quarkonium decays into leptons and photons with 
relativistic kinematics, for both ground states and excited states of 
heavy-quarkonium systems. 

We emphasised with our basically model-independent calculations that either
HQSS holds for radially excited charmonia and there is a not-yet-understood specificity 
in the decay $\eta'_c \to \gamma \gamma$; either HQSS is strongly broken and this hints
at large relativistic corrections at work in such decays . This would in turn
explain why most of available models on the market are unable to give correct 
predictions for both $\eta_c$ and $\eta_c'$ decays.

  Measurements of the two-photon widths for $\eta_{b}$
and  higher excited states  could provide with a test for HQSS
and a determination of the strong coupling constant $\alpha_{s}$ at a scale
around the $\Upsilon$ mass, similarly to what has been done with the $\Upsilon$ leptonic
width in the past.


\begin{thebibliography}{99}

\bibitem{Brambilla}
N.~Brambilla {\it et al.}, {\it Heavy quarkonium physics}, CERN Yellow Report, CERN-2005-005, 
2005  Geneva : CERN, 487 pp 
[arXiv:hep-ph/0412158].

\bibitem{Lansberg:2006dh}
  J.~P.~Lansberg, Int.\ J.\ Mod.\ Phys.\ A {\bf 21} (2006) 3857
  [arXiv:hep-ph/0602091].


\bibitem{Colangelo} For a recent review on open and hidden charm
  spectroscopy, see, e.g. 
 P. Colangelo, F. De Fazio, R. Ferrandes, and  S. Nicotri, Talk given at
the Workshop `` Continuous Advances in QCD 2006'', Minneapolis, USA, May
11-14, 2006, hep-ph/0609240 and other references quoted therein.

\bibitem{Asner}
  D.~M.~Asner {\it et al.}  [CLEO Collaboration],
  Phys.\ Rev.\ Lett.\  {\bf 92} (2004) 142001
  [arXiv:hep-ex/0312058].




\bibitem{Nakazawa} H. Nakazawa, {\it Measurement of $\eta_c(1S,2S)$ in two-photon process at Belle},  talk at the Photon2007 International Conference, July 9-13, 2007,
Paris, France.

\bibitem{eidelman} S. Eidelman, {\it Charmonium studies in $\gamma \gamma$ collisions at Belle}, talk at the 5th International Workshop on Quarkonium, October 17-20, 2007, DESY, Hamburg, Germany.

\bibitem{Ackleh}
 E.~S.~Ackleh and T.~Barnes,
  Phys.\ Rev.\  D {\bf 45}, 232 (1992).

\bibitem{Kim}
  C.~S.~Kim, T.~Lee and G.~L.~Wang,
  Phys.\ Lett.\  B {\bf 606}, 323 (2005)
  [arXiv:hep-ph/0411075].


\bibitem{Ahmady}
  M.~R.~Ahmady and R.~R.~Mendel,
  Phys.\ Rev.\  D {\bf 51}, 141 (1995)
  [arXiv:hep-ph/9401315].

\bibitem{Munz}
  C.~R.~Munz,
  Nucl.\ Phys.\  A {\bf 609}, 364 (1996)
  [arXiv:hep-ph/9601206].

\bibitem{Chao}
  K.~T.~Chao, H.~W.~Huang, J.~H.~Liu and J.~Tang,
  Phys.\ Rev.\  D {\bf 56}, 368 (1997)
  [arXiv:hep-ph/9601381].

\bibitem{Ebert}  
D.~Ebert, R.~N.~Faustov and V.~O.~Galkin,
  Mod.\ Phys.\ Lett.\  A {\bf 18}, 601 (2003)
  [arXiv:hep-ph/0302044].
 

\bibitem{Gerasimov}
  S.~B.~Gerasimov and M.~Majewski, {\it On the relations between two-photon and leptonic widths of low-lying S-wave states of charmonium}, presented at 17th International Baldin Seminar on High Energy Physics Problems: Relativistic Nuclear Physics and Quantum Chromodynamics (ISHEPP 2004), Dubna, Russia, 27 Sep - 1 Oct 2004, 
  arXiv:hep-ph/0504067.

\bibitem{Crater}
  H.~W.~Crater, C.~Y.~Wong and P.~Van Alstine,
  Phys.\ Rev.\ D {\bf 74}, 054028 (2006).
 
\bibitem{Lansberg:2006dw}
  J.~P.~Lansberg and T.~N.~Pham,
  Phys.\ Rev.\ D {\bf 74} (2006) 034001
  [arXiv:hep-ph/0603113].


\bibitem{Lansberg:2006sy}
  J.~P.~Lansberg and T.~N.~Pham,
  Phys.\ Rev.\  D {\bf 75} (2007) 017501
  [arXiv:hep-ph/0609268].

\bibitem{Pham} T.~N.~Pham, Proceedings of the International Workshop on Quantum
Chromodynamics Theory and Experiment, Martina Franca, Valle d'Itria,
Italy, 16-20 Jun 2007, AIP Conf.\ Proc.\  {\bf 964} (2007) 124
  [arXiv:0710.2846 [hep-ph]]


\bibitem{Kuhn}
  J.~H.~K\"uhn, J.~Kaplan and E.~G.~O.~Safiani,
  Nucl.\ Phys.\ B {\bf 157}, 125 (1979).

\bibitem{Guberina}
  B. Guberina,  J.~H.~K\"uhn, R. D.~Peccei, and R. Ruckl,
  Nucl.\ Phys.\ B {\bf 174}, 317 (1980).


\bibitem{Novikov}
  V.~A.~Novikov, L.~B.~Okun, M.~A.~Shifman, A.~I.~Vainshtein, 
  M.~B.~Voloshin and V.~I.~Zakharov,
  Phys.\ Rept.\  {\bf 41}, 1 (1978).

\bibitem{Reinders}
  L.~J.~Reinders, H.~R.~Rubinstein and S.~Yazaki,
  Phys.\ Lett.\ B {\bf 113}, 411 (1982).

\bibitem{Dudek}
  J.~J.~Dudek, R.~G.~Edwards and D.~G.~Richards,
  Phys.\ Rev.\  D {\bf 73} (2006) 074507
  [arXiv:hep-ph/0601137];
  J.~J.~Dudek and R.~G.~Edwards,
  Phys.\ Rev.\ Lett.\  {\bf 97} (2006) 172001
  [arXiv:hep-ph/0607140].

\bibitem{Kwong} 
  W. Kwong, P. B. Mackenzie, R. Rosenfeld and J. L. Rosner,
  Phys.\ Rev.\ D {\bf 37},  3210 (1988).

\bibitem{Pancheri}
  N.~Fabiano and G.~Pancheri,
  Eur.\ Phys.\ J.\  C {\bf 25} (2002) 421
  [arXiv:hep-ph/0204214].


\bibitem{pdg}  
 W.~M.~Yao {\it et al.}  [Particle Data Group],
  J.\ Phys.\ G {\bf 33}, 1 (2006).



\bibitem{Choi}
  S.~K.~Choi {\it et al.}  [BELLE collaboration],
  Phys.\ Rev.\ Lett.\  {\bf 89} (2002) 102001
  [Erratum-ibid.\  {\bf 89} (2002) 129901]
  [arXiv:hep-ex/0206002].




\bibitem{Song}
  Z.~z.~Song, C.~Meng and K.~T.~Chao,
  Eur.\ Phys.\ J.\  C {\bf 36} (2004) 365
  [arXiv:hep-ph/0209257].


\bibitem{Aubert:2005vi}
  B.~Aubert {\it et al.}  [BABAR Collaboration],
  Phys.\ Rev.\ Lett.\  {\bf 96} (2006) 052002
  [arXiv:hep-ex/0510070].


\bibitem{Schuler}
 G.~A.~Schuler, F.~A.~Berends and R.~van Gulik,
  Nucl.\ Phys.\  B {\bf 523} (1998) 423
  [arXiv:hep-ph/9710462].

\bibitem{Lakhina}
  O.~Lakhina and E.~S.~Swanson,
  Phys.\ Rev.\  D {\bf 74} (2006) 014012
  [arXiv:hep-ph/0603164].

\bibitem{Godfrey}
  S.~Godfrey and N.~Isgur,
  Phys.\ Rev.\ D {\bf 32}, 189 (1985).


\bibitem{Fabiano}
  N.~Fabiano,
  Eur.\ Phys.\ J.\  C {\bf 26} (2003) 441
  [arXiv:hep-ph/0209283].


\end{thebibliography}
\end{document}